\documentclass[aps,preprint,nofootinbib]{revtex4}
\usepackage{graphicx}

\begin{document}

\title{B-Meson Wave Function through A Comparative Analysis of the $B\to \pi$, $K$ Transition Form Factors}
\author{Dai-Min Zeng, Xing-Gang Wu\footnote{wuxg@cqu.edu.cn} and Zhen-Yun Fang}
\address{Department of Physics, Chongqing University, Chongqing 400044,
P.R. China}

\begin{abstract}
The properties of the B-meson light-cone wave function up to
next-to-leading order Fock state expansion have been studied through
a comparative study of the $B\to \pi$, $K$ transition form factors
within the $k_T$ factorization approach and the light-cone sum rule
analysis. The transition form factors $F^{B\to\pi}_{+,0,T}$ and
$F^{B\to K}_{+,0,T}$ are carefully re-calculated up to ${\cal
O}(1/m_b^2)$ within the $k_T$ factorization approach in the large
recoil region, in which the main theoretical uncertainties are
discussed. The QCD light-cone sum rule is applicable in the large
and intermediate energy regions, and the QCD light-cone sum rule
results in Ref.\cite{sumrule} are adopted for such a comparative
study. It is found that when the two phenomenological parameters
$\bar\Lambda\in [0.50,0.55]$ and $\delta\in[0.25,0.30]$, the results
of $F^{B\to\pi}_{+,0,T}(Q^2)$ and $F^{B\to K}_{+,0,T}(Q^2)$ from
these two approaches are consistent with each other in the large
recoil energy region. \\



\end{abstract}

\maketitle

The non-perturbative light-cone (LC) wavefunction (WF) of the B
meson plays an important role for making reliable predictions for
exclusive B meson decays. However, the B-meson WF still poses a
major source of uncertainty in the study of B meson decays. Unless
we have known it well and applied it for some precise studies, we
can not definitely say that there is really new physics in the
B-meson decays, e.g. the so called $B\to\pi K$ puzzle
\cite{bpipuzzle} and etc.. Hence, theoretically, it is an important
issue to study on it.

An analytic solution for the leading Fock-state B-meson WF, which is
derived under the Wandzura-Wilczek (WW) approximation \cite{ww} and
with the help of the equation of motion of the light spectator quark
in the B meson, has been given in Refs. \cite{qiao,bwave}. It shows
that the leading Fock-state B-meson WF can be determined uniquely in
analytic form in terms of the "effective mass" ($\bar\Lambda$)
\cite{hqet} of the meson state and its transverse momentum
dependence is just determined through a simple delta function. This
simple model has been frequently used for a leading order estimation
of the B-meson decays. It is argued that when including the
3-particle Fock states' contribution, the transverse momentum
distribution may be expanded to a certain degree other than such a
simple delta function. The contributions from the higher Fock
states' may not be too small, e.g. in Ref.\cite{chli}, the
3-particle contributions are estimated by attaching an extra gluon
to the internal off-shell quark line, and then $(1/m_b)$ power
suppression is readily induced. Recently, a simple model for the
B-meson wave function up to next-to-leading Fock state has been
raised in Ref.\cite{nlob}, where relations between the 2- and 3-
particle wavefunctions derived from the QCD equations of motion and
the heavy quark symmetry \cite{heavyquark}, especially two
constraints derived from the gauge field equation of motion, are
employed. More explicitly, the normalized B-meson wave functions in
the compact parameter $b$-space can be written as \cite{nlob}
\begin{equation}\label{bw1}
\Psi_+(\omega,b)=\frac{\omega}{\omega_0^2}\exp \left(
-\frac{\omega}{\omega_0}\right)\Big(\Gamma[\delta] J_{\delta-1}
[\kappa] +(1-\delta)\Gamma[2-\delta] J_{1-\delta}[\kappa]\Big)\left(
\frac{\kappa}{2} \right)^{1-\delta}
\end{equation}
and
\begin{equation}
\Psi_-(\omega,b)=\frac{1}{\omega_0}\exp \left(
-\frac{\omega}{\omega_0}\right)\Big(\Gamma[\delta] J_{\delta-1}
[\kappa] +(1-\delta)\Gamma[2-\delta] J_{1-\delta}[\kappa]\Big)\left(
\frac{\kappa}{2} \right)^{1-\delta}, \label{bw2}
\end{equation}
with $\omega_0=2\bar\Lambda/3$, $\kappa=\sqrt{\omega
(2\bar\Lambda-\omega)}b $ and $\delta$ is in the range of $(0,1)$.
In the above model, only two typical phenomenological parameters
$\bar\Lambda$ and $\delta$ are introduced. $\bar\Lambda$ stands for
the effective mass of B meson that determines the B-meson's leading
Fock state behavior, while $\delta$ is a typical parameter that
determines the broadness of the B-meson transverse distribution, and
the uncertainty caused by $\delta$ is of order ${\cal O}(1/m_b^2)$.
This solution provides a practical framework for constructing the
B-meson LC WFs and hence is meaningful for phenomenological
applications.

The $B\to \pi$ and $B\to K$ transition form factors
$F^{B\to\pi}_{+,0,T}$ and $F^{B\to K}_{+,0,T}$ provide a good
platform to determine the possible regions for $\bar\Lambda$ and
$\delta$. In the large recoil energy region, the $B\to \pi$, $K$
transition form factors can be studied both under the modified pQCD
factorization approach (or the so-called $k_T$ factorization
approach) \cite{kt,kt2}, the QCD sum rule \cite{svz} and the later
developed QCD light-cone sum rule (LCSR) \cite{lc1}. The properties
of the involving light pseudo-scalar wave functions can be more
precisely determined within the QCD LCSR analysis or the pQCD
calculations from the more sensitive processes like pionic/kaonic
electromagnetic form factors to compare with the experimental data
\cite{em1,em2,em3} or from the lattice calculation \cite{lat1}, so
we shall directly take them to be the ones favored in the
literature, and then the main uncertainties for the present $k_T$
factorization approach come from the B-meson wave function. In fact,
by varying the undetermined parameters of the pionic/kaonic wave
functions within reasonable regions determined in literature
\footnote{In literature, some attempts to derive the properties of
the pionic wave function from $B\to\pi$ within the LCSR approach can
be found in Refs.\cite{bpipball,bpiwu}.}, it can be found that the
main uncertainty of the $B\to \pi$ and $B\to K$ transition form
factors really comes from that of the B-meson wave function
\footnote{Such a discussion for the $B\to K$ vector and scalar form
factors can be found in Ref.\cite{whf}. }. On the other hand, within
the QCD LCSR approach with proper correlator, it has been found that
the main uncertainties in estimation of form factors come from the
pionic and kaonic twist-2 and twist-3 wave functions. A systematic
QCD LCSR calculation of $B\to \pi$, $K$ transition form factors has
been finished in Refs.\cite{sumrule,melic1,melic2,whf1} by including
the one-loop radiative corrections to the pionic/kaonic twist-2 and
twist-3 contributions. So through a comparative study of the form
factors with the $k_{T}$ factorization approach and the QCD LCSR,
one can derive the reasonable regions for the two undetermined
parameters $\bar\Lambda$ and $\delta$ of the B-meson wave function,
which is the main purpose of the present letter.

The $B\to \pi$ and $B\to K$ transition form factors
$F^{B\to\pi}_{+,0,T}$ and $F^{B\to K}_{+,0,T}$ are defined as
follows:
\begin{equation}
\langle P(p)|V^P_\mu| B(p_B)\rangle = \left[(p_B+p)_{\mu}-
\frac{M_B^2-M_P^2}{q^2}q_{\mu}\right] F_+^{B\to K}(q^2)+
\frac{M_B^2-M_P^2}{q^2} q_{\mu}F_0^{B\to P}(q^2)
\end{equation}
and
\begin{equation}
\langle P(p)|J^{P,\sigma}_{\mu}| B(p_B)\rangle = i\frac{F_T^{B\to
P}(q^2)}{M_B +M_P} \left[q^2 (p_B+p)_{\mu}- (M_B^2-M_P^2) q_{\mu}
\right] ,
\end{equation}
where $P$ stands for the pseudo-scalar meson $\pi$ or $K$
respectively, the momentum transfer $q=p_{B}-p$, the vector currents
$V^{\pi}_{\mu}=\bar{u}\gamma_{\mu}b$ and
$V^{K}_{\mu}=\bar{s}\gamma_{\mu}b$, the tensor currents
$J^{\pi,\sigma}_{\mu}=q^{\mu}\bar{d}\sigma_{\mu\nu}b$ and
$J^{K,\sigma}_{\mu}=q^{\mu}\bar{s}\sigma_{\mu\nu}b$.

Within the $k_T$ factorization approach, the $B\to P$ transition
form factors are dominated by a single gluon exchange in the lowest
order. Following the same procedure as described in
Ref.\cite{huang}, we obtain all the mentioned transition form
factors in the transverse configuration ${\bf b}$-space up to order
${\cal O}(1/m_b^2)$, i.e.
\begin{eqnarray}
F_+^{B\to P}(q^2) &=& \frac{\pi C_F}{N_c} f_{P}f_B M_B^2\int d\xi
dx\int b_B db_B ~ b_P db_P ~ \alpha_s(t) \times\exp(-S(x,\xi,b_P,b_B;t)) \nonumber\\
&\times& S_t(x)S_t(\xi)\Bigg \{ \Bigg [ \Psi_P(x, b_P)\left (
(x\eta+1)\Psi_B(\xi, b_B)-\bar\Psi_B(\xi, b_B) \right)\nonumber \\
&+& \frac{m_0^p}{M_B}\Psi_p(x, b_P)\cdot\left( (1-2x)\Psi_B
(\xi,b_B) +\left(x+\frac{1}{\eta}-1\right)\bar\Psi_B(\xi,b_B)\right)\nonumber\\
&-&\frac{m_0^p}{M_B}\frac{\Psi'_\sigma(x,b_P)}{6}
\cdot\left(\left(1+2x-\frac{2}
{\eta}\right)\Psi_B(\xi,b_B)-(1+x-\frac{1}
{\eta})\bar\Psi_B(\xi,b_B) \right) \nonumber\\
&+& \frac{m_0^p}{M_B}\Psi_\sigma(x,b_P)\left( \Psi_B(\xi,b_B)
-\frac{\bar\Psi_B(\xi,b_B)}{2}\right)\Bigg
]h_1(x,\xi,b_P,b_B)\nonumber\\
&-& (1+\eta+x\eta) \frac{m_0^p}{M_B}\frac{\Psi_\sigma(x,b_P)}{6}
[M_B\Delta(\xi,b_B)]h_2(x,\xi,b_P,b_B)\nonumber \\
&+& \Bigg [ \Psi_P(x, b_p)\left ( -\xi\bar\eta
\Psi_B(\xi,b_B)+\frac{\Delta(\xi,b_B)}{M_B}
\right )+2\frac{m_0^p}{M_B}\Psi_p(x,b_P)\cdot\nonumber \\
&& \left( (1-\xi)\Psi_B(\xi,b_B)+\xi(1-\frac{1}
{\eta})\bar\Psi_B(\xi,b_B) +2\frac{\Delta(\xi,b_B)} {M_B}\right )
\Bigg ] h_1(\xi,x,b_B,b_P) \Bigg \}, \label{fbc+}
\end{eqnarray}
\begin{eqnarray}
F_0^{B\to P}(q^2) &=& \frac{\pi C_F}{N_c} f_P f_B M_B^2\int d\xi
dx\int b_Bdb_B~ b_P db_P~ \alpha_s(t) \times \exp(-S(x,\xi,b_P,b_B;t))\nonumber\\
&\times& S_t(x)S_t(\xi) \Bigg \{ \Bigg [ \Psi_P(x, b_P)\eta\left (
(x\eta+1)\Psi_B(\xi, b_B)-\bar\Psi_B(\xi, b_B) \right ) \nonumber \\
&+&\frac{m_0^p}{M_B}\Psi_p(x, b_P) \big((2-\eta-2x\eta)
\Psi_B(\xi,b_B)-(1-\eta-x\eta)\bar\Psi_B(\xi,b_B)\big) \nonumber \\
&-&\frac{m_0^p}{M_B}\frac{\Psi'_\sigma(x,b_P)}{6}\cdot\big(
\eta(2x-1)\Psi_B(\xi,b_B)-(1+x\eta-\eta)\bar\Psi_B(\xi,b_B) \big) \nonumber\\
&+&\eta\frac{m_0^p}{M_B} \Psi_\sigma(x,b_P) \left(\Psi_B(\xi,b_B)
-\frac{\bar\Psi_B(\xi,b_B)}{2}\right)\Bigg ] h_1(x,\xi,b_P,b_B) \nonumber \\
&-& [3-\eta-x\eta]\frac{m_0^p}{M_B} \frac{\Psi_\sigma(x,b_P)}{6}
[M_B\Delta(\xi,b_B)]h_2(x,\xi,b_P,b_B)\nonumber \\
&+& \Bigg [ \Psi_P(x, b_P)\eta\left ( \xi\bar\eta \Psi_B(\xi,b_B)
+\frac{\Delta(\xi,b_B)}{M_B} \right )\nonumber \\
&+& 2\frac{m_0^p}{M_B}\Psi_p(x,b_P)\cdot\Big(
(\eta(1+\xi)-2\xi)\Psi_B(\xi,b_B) -(\eta\xi-\xi)\bar\Psi_B(\xi,b_B)
\nonumber\\
&+& 2(2-\eta) \frac{\Delta(\xi,b_B)}{M_B}\Big) \Bigg ]
h_1(\xi,x,b_B,b_P) \Bigg \}\label{fbc0}
\end{eqnarray}
and
\begin{eqnarray}
F_T^{B\to P}(q^2) &=& \frac{\pi C_F}{N_c} f_P f_B M_B^2\int d\xi
dx\int b_B db_B~ b_P db_P~ \alpha_s(t)
\times\exp[-S(x,\xi,b_P,b_B;t)] \nonumber\\
&&\times S_t(x)S_t(\xi)\Bigg \{ \Bigg [ \Psi_P(x, b_P)\left (
\Psi_B(\xi, b_B)-\bar\Psi_B(\xi, b_B) \right )+
\frac{m_0^p}{M_B}\Psi_p(x, b_P )\cdot\nonumber \\
&&\left(\frac{1}{\eta}\bar\Psi_B(\xi,b_B) -x\Psi_B(\xi,b_B)\right)
+\frac{m_0^p}{M_B}\frac{\Psi'_\sigma(x,b_P)}{6}
\left(\frac{x\eta+2}{\eta}\Psi_B(\xi,b_B)\right.\nonumber\\
&&\left. -\frac{1}{\eta}\bar\Psi_B(\xi,b_B)
\right)+\frac{m_0^p}{M_B} \frac{\Psi_\sigma(x,b_P
)}{6}\Psi_B(\xi,b_B)\Bigg ] h_1(x,\xi,b_P,b_B)- \frac{m_0^p}{M_B}
\frac{\Psi_\sigma(x,b_P )}{6}\nonumber \\
&& [M_B\Delta(\xi,b_P)]h_2(x,\xi,b_P,b_B) +\Bigg [ \Psi_P(x,
b_P )\left (\frac{\Delta(\xi,b_B)}{M_B}-\xi\Psi_B(\xi,b_B)\right )+\nonumber\\
&& 2\frac{m_0^p}{M_B}\Psi_p(x,b_P )\left(
\Psi_B(\xi,b_B)-\frac{\xi}{\eta}\bar\Psi_B(\xi,b_B)\right ) \Bigg ]
h_1(\xi,x,b_B,b_P ) \Bigg \}, \label{fbcT}
\end{eqnarray}
where the integration over the azimuth angles have been implicitly
done, the transverse momentum dependence for both the hard
scattering part and the non-perturbative wave functions, the Sudakov
effects and the threshold effects are included to give a consistent
analysis of the form factors up to ${\cal O} (1/m^2_b)$. And the two
introduced functions
\begin{eqnarray}
h_1(x,\xi,b_P,b_B)&=&K_0(\sqrt{\xi x\eta}~M_B b_B) \Bigg [
\theta(b_B-b_P)I_0(\sqrt{x\eta}~M_Bb_P)K_0(\sqrt{x\eta}~M_B b_B) \nonumber \\
&&+\theta(b_P-b_B)I_0(\sqrt{x\eta}~M_Bb_B)K_0(\sqrt{x\eta}~M_B b_P)
\Bigg ]
\end{eqnarray}
and
\begin{eqnarray}
h_2(x,\xi,b_P,b_B)&=&\frac{b_B}{2\sqrt{\xi x\eta}M_B}K_{1}(\sqrt{\xi
x\eta}~M_B b_B)\Bigg [ \theta(b_B-b_P)I_0(\sqrt{x\eta}~M_Bb_P)
K_0(\sqrt{x\eta}~M_B b_B) \nonumber \\
&&+\theta(b_P-b_B)I_0(\sqrt{x\eta}~M_Bb_B)K_0(\sqrt{x\eta}~M_B b_P)
\Bigg ] ,
\end{eqnarray}
where the functions $I_i$ ($K_i$) are the modified Bessel functions
of the first (second) kind with the $i$-{\it th} order. Implicitly,
we have set
\begin{equation}
\Psi_B=\Psi_B^{+}\ ,\;\;\; \bar{\Psi}_B=\Psi_B^{+}-\Psi_B^-\ ,\;\;\;
\Delta(\xi, b_B) =M_B \int_0^{\xi} d\xi' [\Psi^-_B(\xi',b_B)
-\Psi^+_B(\xi',b_B)] ,
\end{equation}
where B-meson wave functions $\Psi_B^{+}$ and $\Psi_B^-$ are taken
as Eqs.(\ref{bw1},\ref{bw2}) that include 3-particle Fock states'
contributions. It can be found that the contributions from
$\bar{\Psi}_B$ is rightly power suppressed to that of $\Psi_B$. Such
a definition for $\Psi_B$ and $\bar{\Psi}_B$ is often adopted in the
literature to simplify the calculation, since the the contribution
from $\bar\Psi_B$ can be safely neglected for the leading order
estimation \footnote{Another definition is also adopted in
literature, $\Psi_B=(\Psi_B^{+}+\Psi_B^-)/2$ and
$\bar{\Psi}_B=(\Psi_B^{+}-\Psi_B^-)/2$, however in this definition
$\Psi_B$ and $\bar\Psi_B$ should be treated on the equal footing as
pointed out in Ref.\cite{huang}.}. The factor
$\exp(-S(x,\xi,b_P,b_B;t))$ contains the Sudakov logarithmic
corrections and the renormalization group evolution effects of both
the wave functions and the hard scattering amplitude,
\begin{equation}
S(x,\xi,b_P,b_B;t)=
\left[s(x,b_P,M_b)+s(\bar{x},b_P,M_b)+s(\xi,b_B,M_b)
-\frac{1}{\beta_{1}}\ln\frac{\hat{t}}{\hat{b}_\pi}
-\frac{1}{\beta_{1}}\ln\frac{\hat{t}}{\hat{b}_B} \right],
\end{equation}
where ${\hat t}={\rm ln}(t/\Lambda_{QCD})$, ${\hat b}_B ={\rm
ln}(1/b_B\Lambda_{QCD})$, ${\hat b}_\pi ={\rm
ln}(1/b_P\Lambda_{QCD}) $ and $s(x,b,Q)$ is the Sudakov exponent
factor, whose explicit form up to next-to-leading log approximation
can be found in Ref.\cite{liyu}. $S_t(x)$ and $S_t(\xi)$ come from
the threshold resummation effects and here we take a simple
parametrization proposed in Refs.\cite{li1,kls},
\begin{equation}
S_t(x)=\frac{2^{1+2c}\Gamma(3/2+c)}{\sqrt{\pi}\Gamma(1+c)}
[x(1-x)]^c\;,
\end{equation}
where the parameter $c$ is determined around $0.3$ for the present
case. The hard scale $t$ in $\alpha_s(t)$ and the Sudakov form
factor might be varied for the different hard scattering parts and
here we need two $t_i$, which can be chosen as the largest scale of
the virtualities of internal particles \cite{li1,lucai},
\begin{equation}
t_1={\rm MAX}\left(\sqrt{x\eta}M_B,1/b_P ,1/b_B\right),\; t_2={\rm
MAX} \left(\sqrt{\xi\eta}M_B,1/b_P ,1/b_B\right).
\end{equation}
The Fourier transformation for the transverse part of the wave
function is defined as
\begin{equation}\label{fourier}
\Psi(x,\mathbf{b})=\int_{|\mathbf{\mathbf{k}}|<1/b}
d^2\mathbf{k}_\perp\exp\left(-i\mathbf{k}_\perp
\cdot\mathbf{b}\right)\Psi(x,\mathbf{k}_\perp),
\end{equation}
where $\Psi$ stands for $\Psi_P$, $\Psi_p$, $\Psi_\sigma$, $\Psi_B$,
$\bar\Psi_B$ and $\Delta$, respectively. The upper edge of the
integration $|\mathbf{k}_\perp|<1/b$ is necessary to ensure that the
wave function is soft enough \cite{huang2}. And we take the
phenomenological parameter, which is a scale characterized by the
chiral perturbation theory, $m^{p}_{0}\simeq 1.30$ GeV for pion
\cite{pimp0} and $m^{p}_{0}\simeq 1.70GeV$ for kaon \cite{lucai}
respectively. For the twist-3 $\Psi_{p}(x,\mathbf{k}_\perp)$ and
$\Psi_{\sigma}(x,\mathbf{k}_\perp)$, we take them to be the ones
constructed in Ref.\cite{huang} for the pionic case and
Ref.\cite{whf} for the kaonic case respectively. As for the twist-2
pion and kaon WFs, they can be constructed based on their first two
Gegenbauer moments and the BHL prescription \cite{bhl}, i.e.
\begin{equation}
\Psi_{\pi}(x,\mathbf{k}_\perp) = [1+B_\pi C^{3/2}_2(2x-1)+C_\pi
C^{3/2}_4(2x-1)]\frac{A_\pi}{x(1-x)} \exp \left[-\beta_\pi^2
\left(\frac{\mathbf{k}_\perp^2+m_q^2}{x(1-x)}\right)\right],
\end{equation}
and
\begin{equation}
\Psi_{K}(x,\mathbf{k}_\perp) = [1+B_K C^{3/2}_1(2x-1)+C_K
C^{3/2}_2(2x-1)]\frac{A_K}{x(1-x)} \exp \left[-\beta_K^2
\left(\frac{\mathbf{k}_\perp^2+m_q^2}{x}+
\frac{\mathbf{k}_\perp^2+m_s^2} {1-x}\right)\right],
\end{equation}
where $q=u,\; d$, $C^{3/2}_{1,2}(1-2x)$ are Gegenbauer polynomials.
The constitute quark masses are set to be: $m_q=0.30{\rm GeV}$ and
$m_s=0.45{\rm GeV}$. The four undetermined parameters can be
determined by the first two Gegenbauer moments $a^\pi_2$ and
$a^\pi_4$ (or $a^K_1$ and $a^K_2$), the normalization condition and
the constraint $\langle \mathbf{k}_\perp^2 \rangle^{1/2}_K \approx
\langle \mathbf{k}_\perp^2 \rangle^{1/2}_\pi=0.350{\rm GeV}$
\cite{gh}, where the average value of the transverse momentum square
is defined as
\begin{displaymath}
\langle \mathbf{k}_\perp^2 \rangle^{1/2}_{\pi,K}=\frac{\int dx
d^2\mathbf{k}_\perp |\mathbf{k}_\perp^2| |\Psi_{\pi,K}(x,{\bf
k}_{\perp})|^2} {\int dx d^2\mathbf{k}_\perp |\Psi_{\pi,K}(x,{\bf
k}_{\perp})|^2} .
\end{displaymath}

As a comparison, within the QCD LCSR approach with proper choosing
correlator, it has been found that the main uncertainties in
estimation of those form factors come from the pionic and kaonic
twist-2 and twist-3 wave functions, especially from the twist-3 wave
function $\Psi_p$ whose distribution amplitude (DA) $\phi_p$ has the
asymptotic behavior $\phi_p(x)|_{q^2\to\infty} \to 1$ for both the
pionic and kaonic cases. Two typical ways have been adopted to
suppress the uncertainty caused by the twist-3 WFs. One way is
raised by Ref.\cite{huangsr1}, i.e. an improved LCSR with proper
chiral current was adopted to eliminate the contributions from the
most uncertain pionic and kaonic twist-3 wave functions and to
enhance the reliability of the LCSR calculations
\cite{srpi,hlk,hwk}. The other way is to do a systematic QCD LCSR
calculation of $B\to \pi$, $K$ transition form factors by including
the one-loop radiative corrections to both the pionic/kaonic twist-2
contributions and the twist-3 contributions
\cite{sumrule,melic1,melic2}. A comparison of these two approaches
to improve the LCSR estimation has been done in Ref.\cite{hwk},
which shows that these two treatments are equivalent to each other,
at least for $F^{B \to \pi}_{+}(q^2)$ and $F^{B\to K}_{+}(q^2)$.
Here we shall adopt the LCSR results of Ref.\cite{sumrule} to do our
discussion, where $F^{B\to\pi}_{+,0,T}$ and $F^{B\to K}_{+,0,T}$
have been parameterized in the following form \cite{sumrule}
\begin{equation}\label{sumruleapp}
F^{B\to P}_{+,0,T}(q^2)=f^{as}(q^2)+a^P_1(\mu_0)
f^{a^P_1}(q^2)+a^P_2(\mu_0) f^{a^P_2}(q^2)+a^P_4(\mu_0)
f^{a^P_4}(q^2),
\end{equation}
where $P$ stands for $\pi$ or $K$, $f^{as}$ contains the
contributions to the form factor from the asymptotic DA and all
higher-twist effects from three-particle quark-quark-gluon matrix
elements, $f^{a^P_1,a^P_2,a^P_4}$ contains the contribution from the
higher Gegenbauer term of DA that is proportional to $a^P_1$,
$a^P_2$ and $a^P_4$ respectively. $\mu_0$ is the factorization scale
which separates long-distance physics (distribution amplitudes) from
the short-distance physics (hard-scattering amplitudes). The
explicit expressions of $f^{as,a^P_1,a^P_2,a^P_4}$ can be found in
Table V and Table IX of Ref.\cite{sumrule}. Since the form factors
have been split into contributions from different Gegenbauer
moments, and the uncertainties other than the Gegenbauer moment
itself have been absorbed into the uncertainty of the functions
$f^{as}$ and $f^{a^P_1,a^P_2,a^P_4}$, then one can conveniently
obtain the LCSRs with various Gegenbauer moments with the help of
Eq.(\ref{sumruleapp}). Eq.(\ref{sumruleapp}) allows one to use the
possible newly developed pionic or kaonic twist-2 DA Gegenbauer
moments to do the discussion, and it is the reason why we chose the
LCSRs of Ref.\cite{sumrule} to do our comparison.

In doing the numerical calculations, we take
$\Lambda^{(n_f=4)}_{\over{MS}}=250MeV$ and
\begin{eqnarray}
f_\pi=130.7\pm0.1\pm0.36 MeV,\;\; f_K=159.8\pm1.4\pm0.44 MeV ,
\end{eqnarray}
where the decay constants $f_\pi$ and $f_K$ are taken from
Ref.\cite{pdg}. As for $f_B$, it can be calculated from the QCD sum
rules and the lattice QCD. Here we fix its value to be $190 MeV$,
i.e. the center value derived from the lattice QCD \cite{fblat},
which is consistent with the Belle experiment
$f_B=229^{+36+34}_{-31-37} MeV$ \cite{belle}. The change of $f_B$
can influence the final results on the transition form factors a lot
and a more precise $f_B$ shall be helpful to improve the precision
of our present estimation, e.g. it can be found that a variation of
$f_B$ by $10\%$ shall bring less than $3\%$ extra uncertainty to the
allowable range of $\bar{\Lambda}$ and $\delta$. A more detailed
discussion of $f_B$ with the LCSRs up to next-to-leading order shall
be presented elsewhere \cite{fbnlo}.

\begin{figure}
\centering
\includegraphics[width=0.3\textwidth]{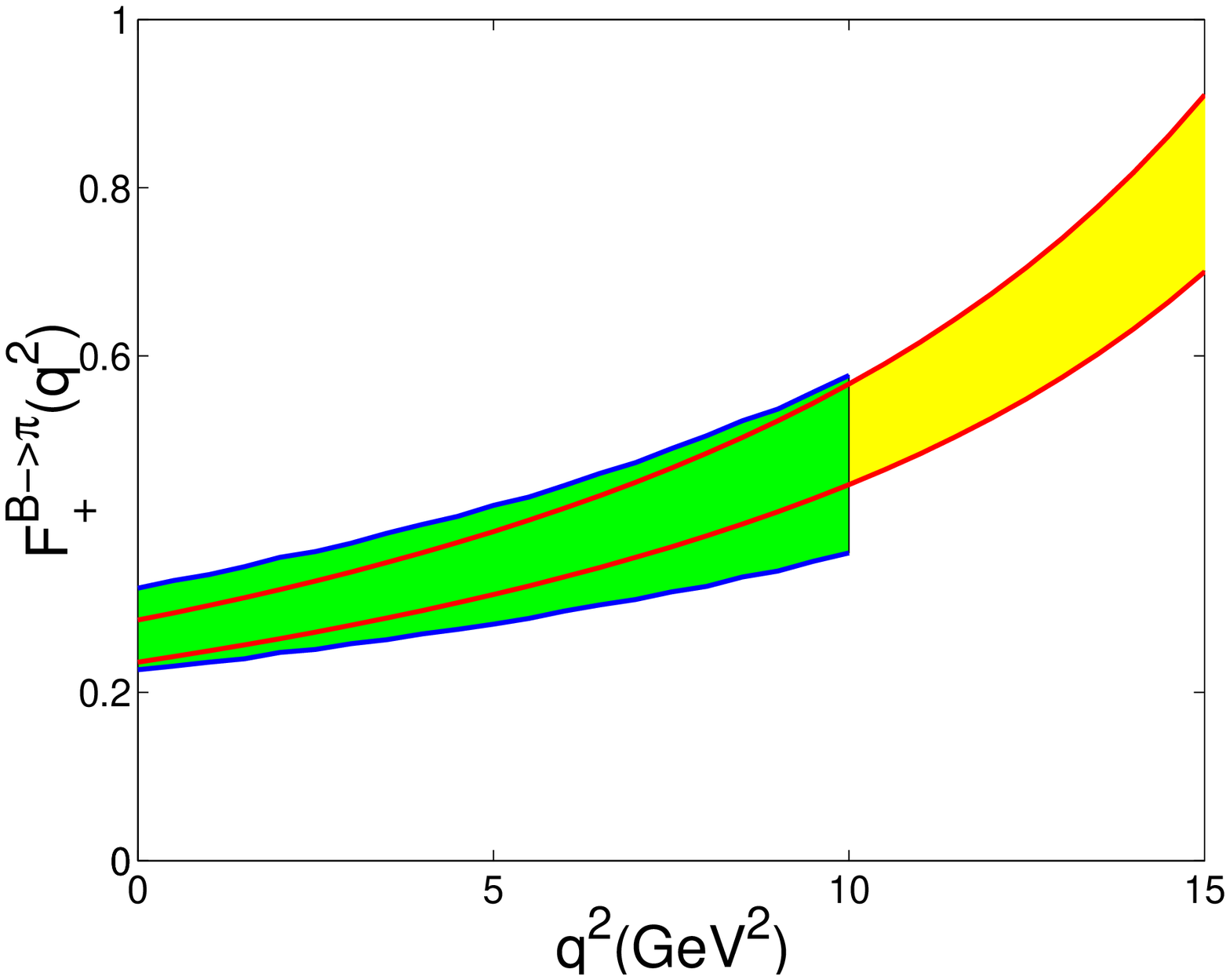} %
\includegraphics[width=0.3\textwidth]{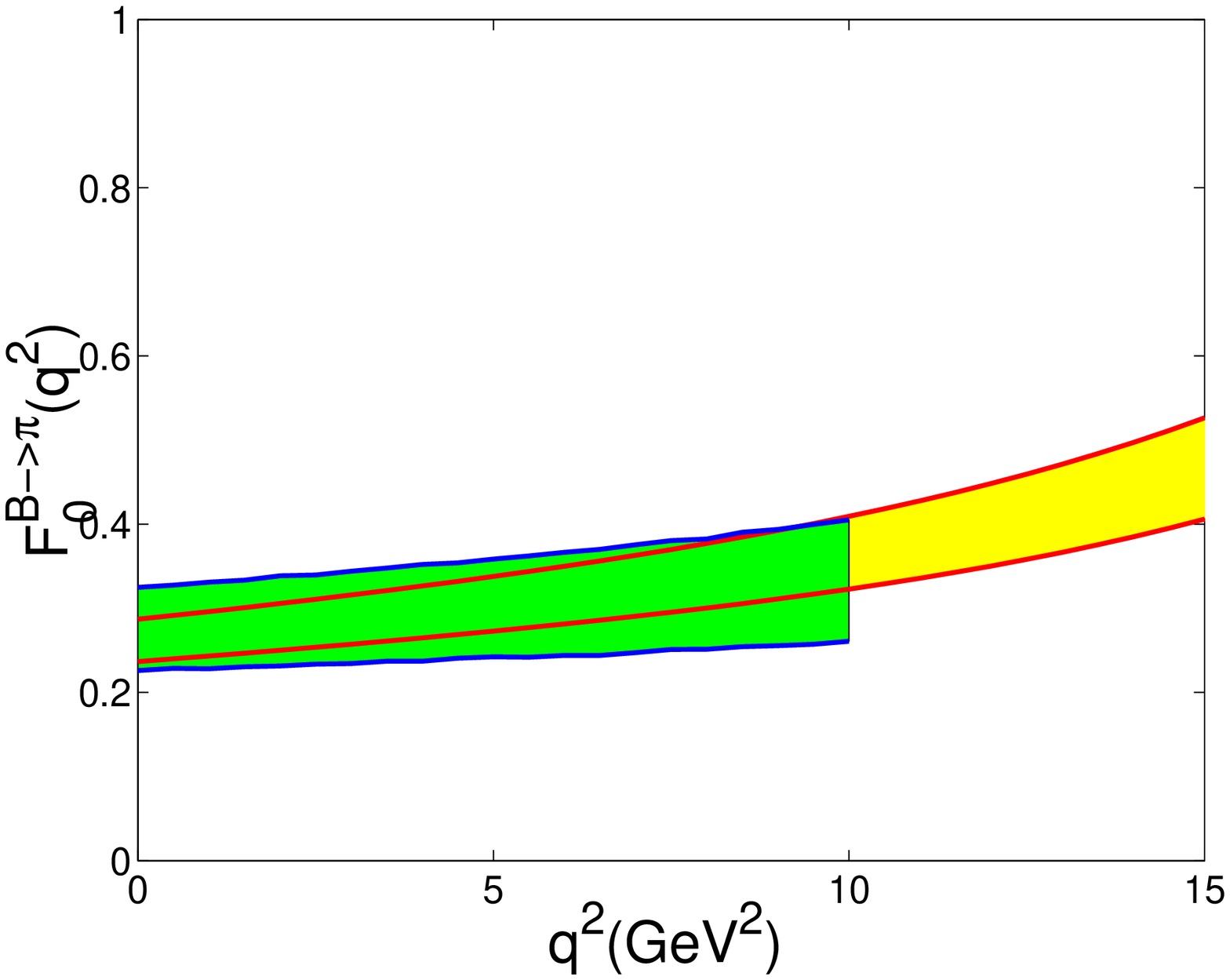} %
\includegraphics[width=0.3\textwidth]{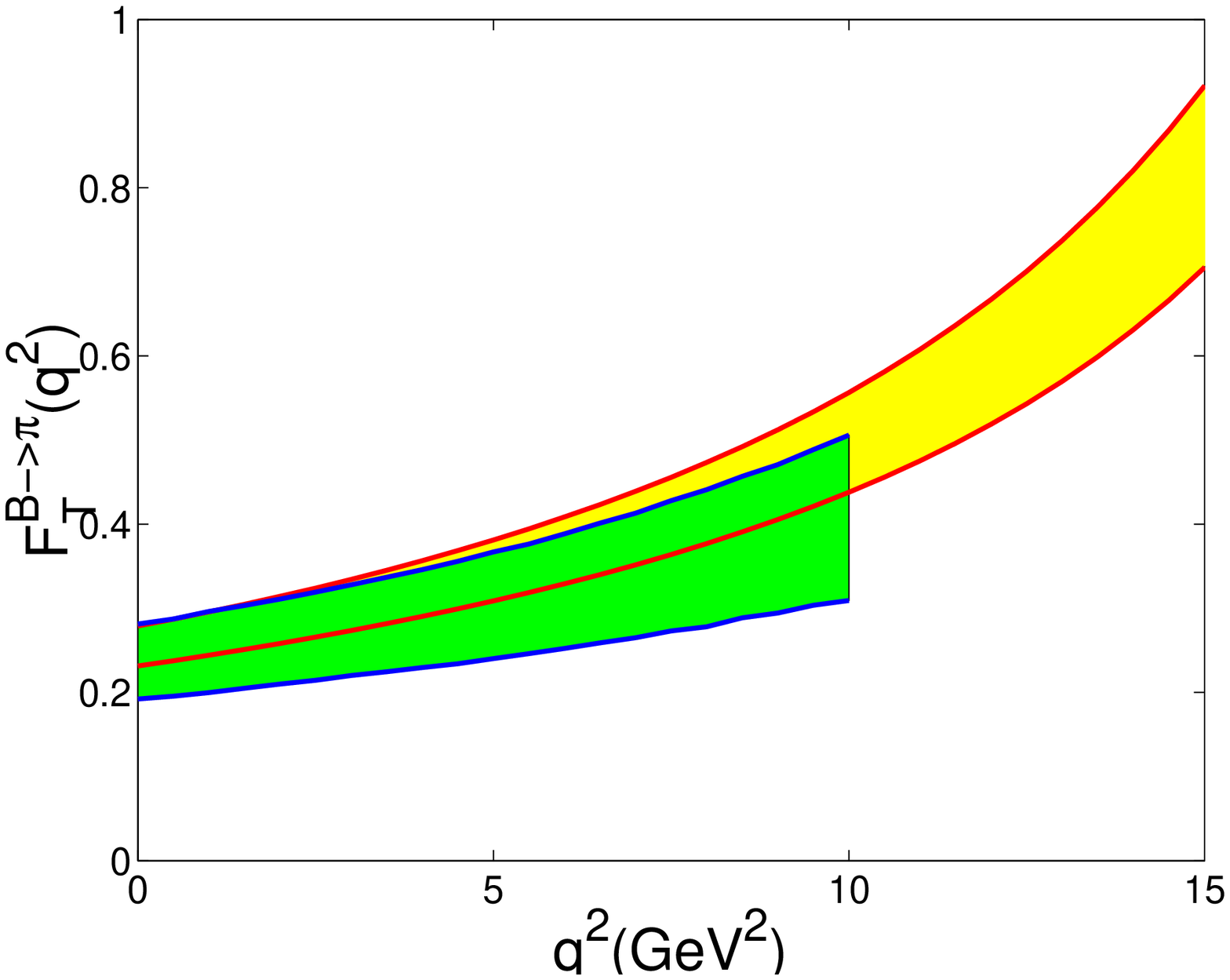}
\caption{Comparative results of $F_{+,0,T}^{B\to \pi}(q^2)$ within
the $k_T$ factorization approach and the QCD LCSR, where the fuscous
shaded band stands for $k_T$ factorization results and the grey band
stands for the QCD LCSR results \cite{sumrule}. The upper edge of
the fuscous shaded band is for $\bar\Lambda=0.50GeV$ and
$\delta=0.30$, and the lower edge of the fuscous shaded band is for
$\bar\Lambda=0.55GeV$ and $\delta=0.25$. } \label{btopi}
\end{figure}

The $B\to \pi$ transition form factors $F^{B\to\pi}_{+,0,T}$ are
shown in Fig.(\ref{btopi}), where $a_2^\pi$ and $a_4^\pi$ are within
the region determined by the two suggested constraints
\cite{sumrule}: $a_2^{\pi}(1 \mbox{GeV}) +a_4^{\pi}(1
\mbox{GeV})=0.1\pm 0.1$ \cite{Bakulev} and $-\frac{9}{4}a_2^{\pi}(1
\mbox{GeV}) +\frac{45}{16}a_4^{\pi}(1 \mbox{GeV})+\frac{3}{2}=1.2\pm
0.3$ \cite{braun0}. The results of the $k_T$ factorization with
$\bar\Lambda\in [0.50,0.55]$ and $\delta\in[0.25,0.30]$ are shown by
a fuscous shaded band with $0\leq q^2\leq 10GeV^2$, and the LCSR
results with its $12\%$ uncertainty \cite{sumrule} are shown by a
grey band with $0\leq q^2\leq 15GeV^2$. It can be found that the
form factors $F^{B\to\pi}_{+,0,T}$ decrease with the increment of
$\bar\Lambda$ and increase with the increment of $\delta$. The upper
edge of the fuscous shaded band is for $\bar\Lambda=0.50GeV$ and
$\delta=0.30$, and the lower edge of the fuscous shaded band is for
$\bar\Lambda=0.55GeV$ and $\delta=0.25$. One may observe that $k_T$
factorization results of $F^{B\to\pi}_{+,0,T}$ can agree with that
of the QCD LCSR at small value of $q^{2}$ with proper values for
$\bar\Lambda$ and $\delta$. And the best fit of all the three form
factors within these two approaches are obtained for
$\bar\Lambda\simeq 0.525GeV$ and $\delta\simeq0.275$.

\begin{figure}
\centering
\includegraphics[width=0.3\textwidth]{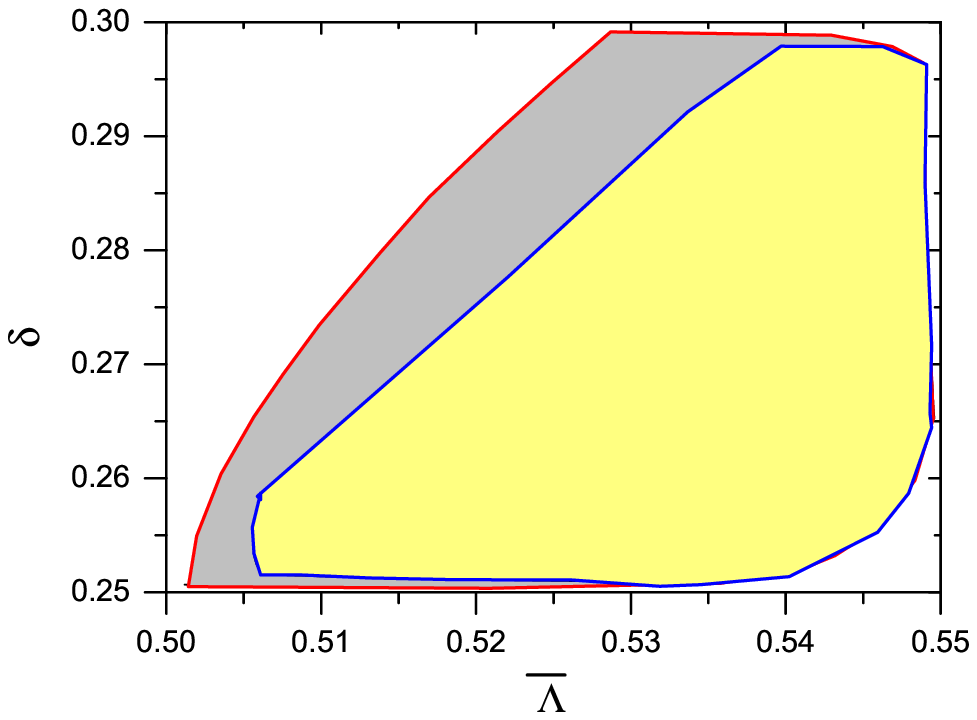} %
\includegraphics[width=0.3\textwidth]{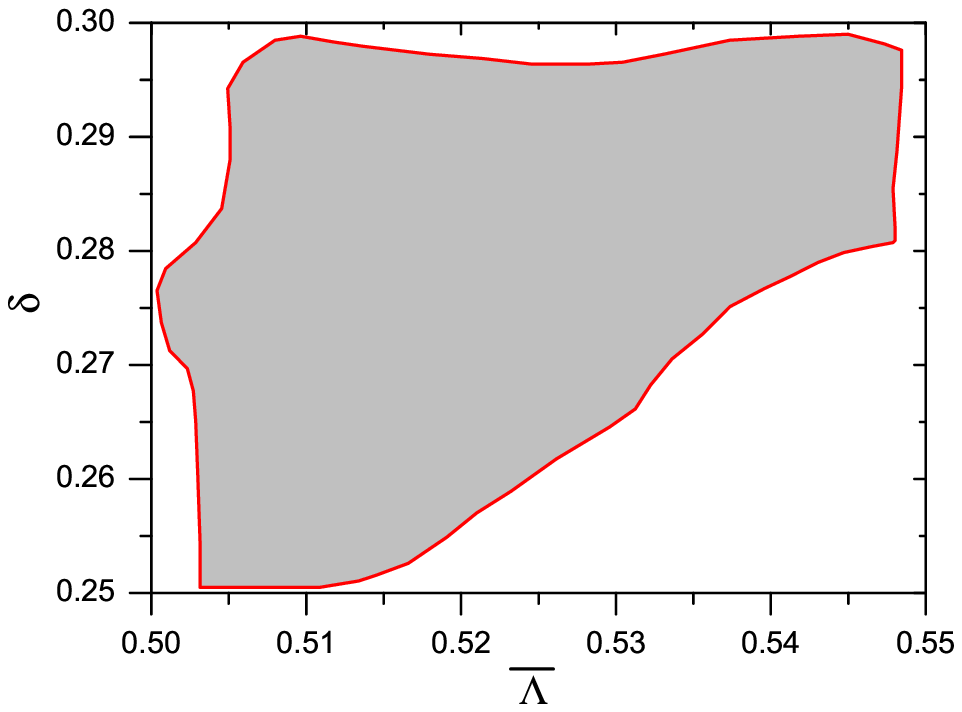} %
\includegraphics[width=0.3\textwidth]{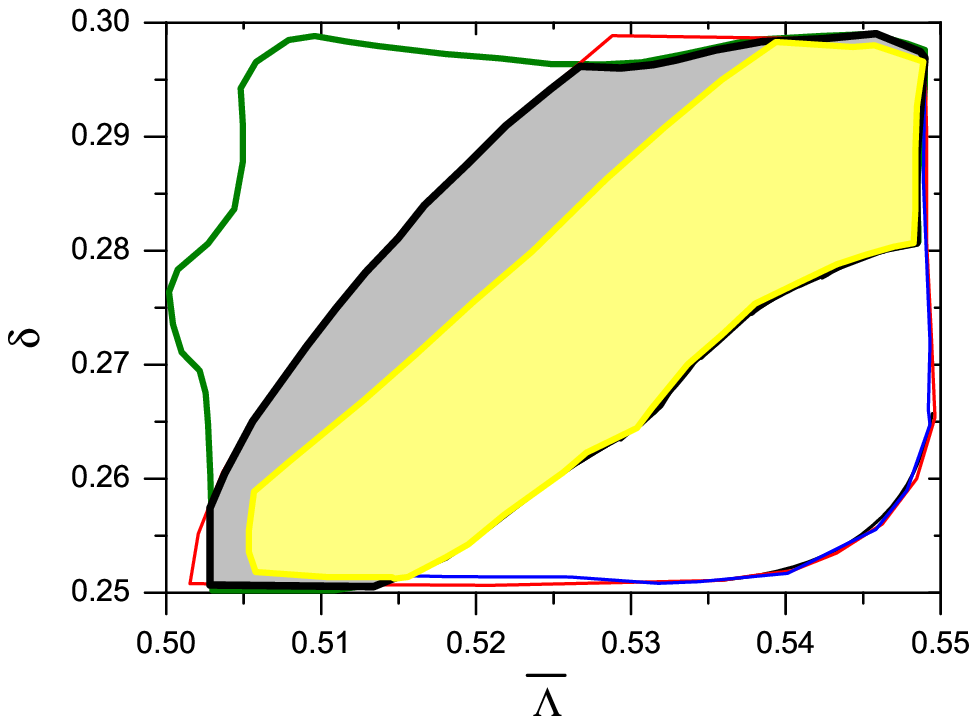}
\caption{Allowable regions for $\bar\Lambda$ and $\delta$ from
$B\to\pi$ form factors $F^{B\to\pi}_{+,0,T}(0)$, which are shown by
fuscous shaded bands respectively. Left diagram stands for the
constraints from $F^{B\to\pi}_{+,0}(0)$, Middle one is for
$F^{B\to\pi}_{T}(0)$ and Right one is the combined results of the
two, where the fainter band is for $F^{B\to\pi}_{+}(0)=0.26\pm0.02$
\cite{pball0}. } \label{pirange}
\end{figure}

At $q^2=0$, the QCD LCSR gives \cite{sumrule}:
$F^{B\to\pi}_{+,0}(0)=0.258\pm 0.031$ and
$F^{B\to\pi}_{T}(0)=0.253\pm 0.028$. If requiring the $k_T$
factorization results for $F^{B\to\pi}_{+,0,T}(0)$ to be consistent
with that of LCSR, one can obtain the possible ranges for
$\bar\Lambda$ and $\delta$, i.e. $\bar\Lambda\in[0.50,0.55]GeV$ and
$\delta\in[0.25,0.30]$, and furthermore, $\bar\Lambda$ and $\delta$
should be correlated in a way as shown in Fig.(\ref{pirange}). In
Fig.(\ref{pirange}), the left diagram stands for the constraints
from $F^{B\to\pi}_{+,0}(0)$, the middle one is for
$F^{B\to\pi}_{T}(0)$ and the right one is the combined results from
$F^{B\to\pi}_{+,0}(0)$ and $F^{B\to\pi}_{T}(0)$. Recently, a nearly
model-independent analysis for $F^{B\to\pi}_{+}(q^2)$ based on the
BaBar experimental data on $B\to\pi l\nu$ has been given in
Ref.\cite{pball0}, which shows $F^{B\to\pi}_{+}(0)=0.26\pm0.02$. If
requiring the $k_T$ factorization results for $F^{B\to\pi}_{+}(0)$
to be within this smaller region, we can obtain a more stringent
constraints for $\bar\Lambda$ and $\delta$ as is shown by the
fainter band of Fig.(\ref{pirange}). Note, all the contours are
obtained by sampling 10,000 points for $F^{B\to\pi}_{+,0,T}(0)$ to
be within the allowable region respectively.

\begin{figure}
\centering
\includegraphics[width=0.3\textwidth]{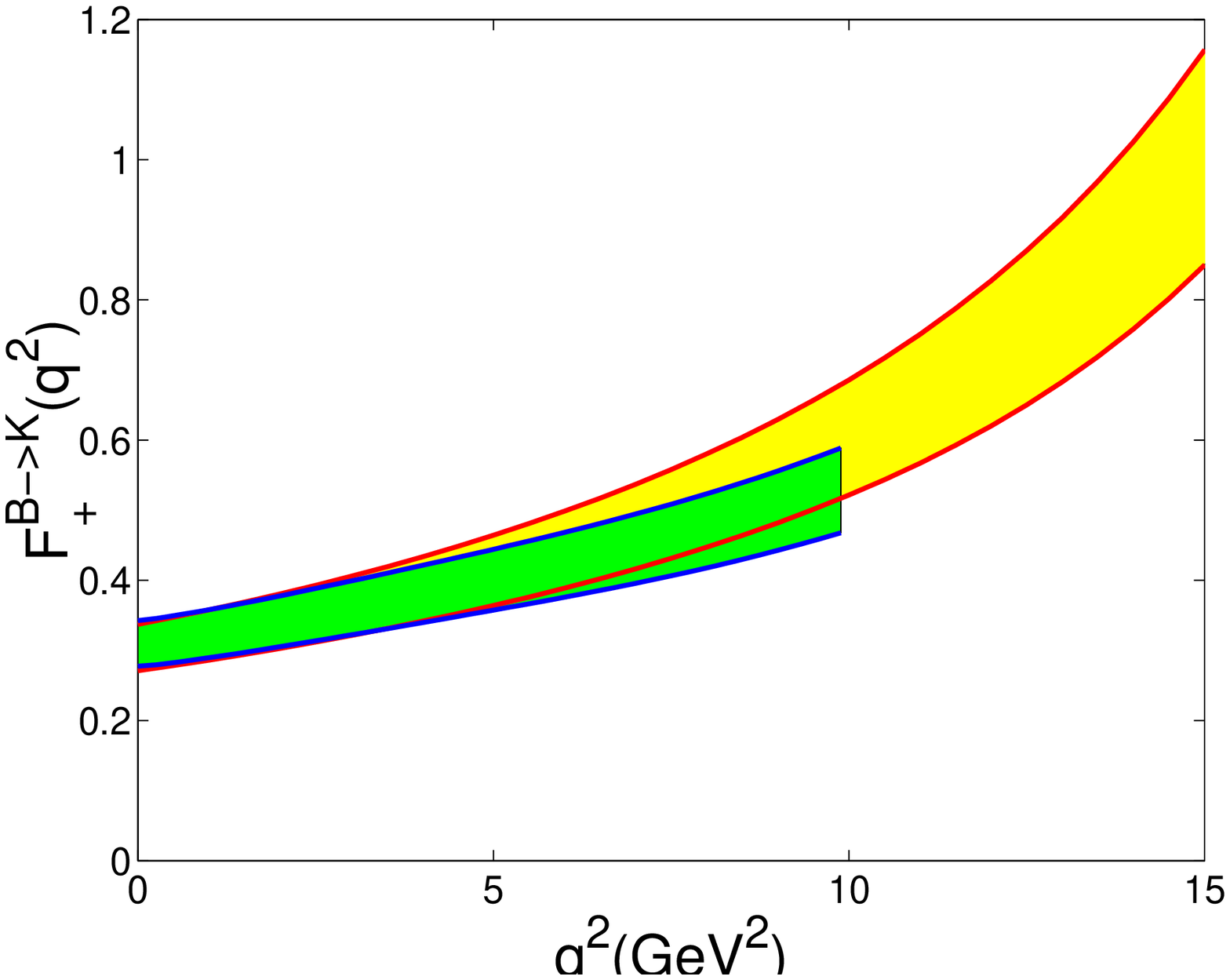} %
\includegraphics[width=0.3\textwidth]{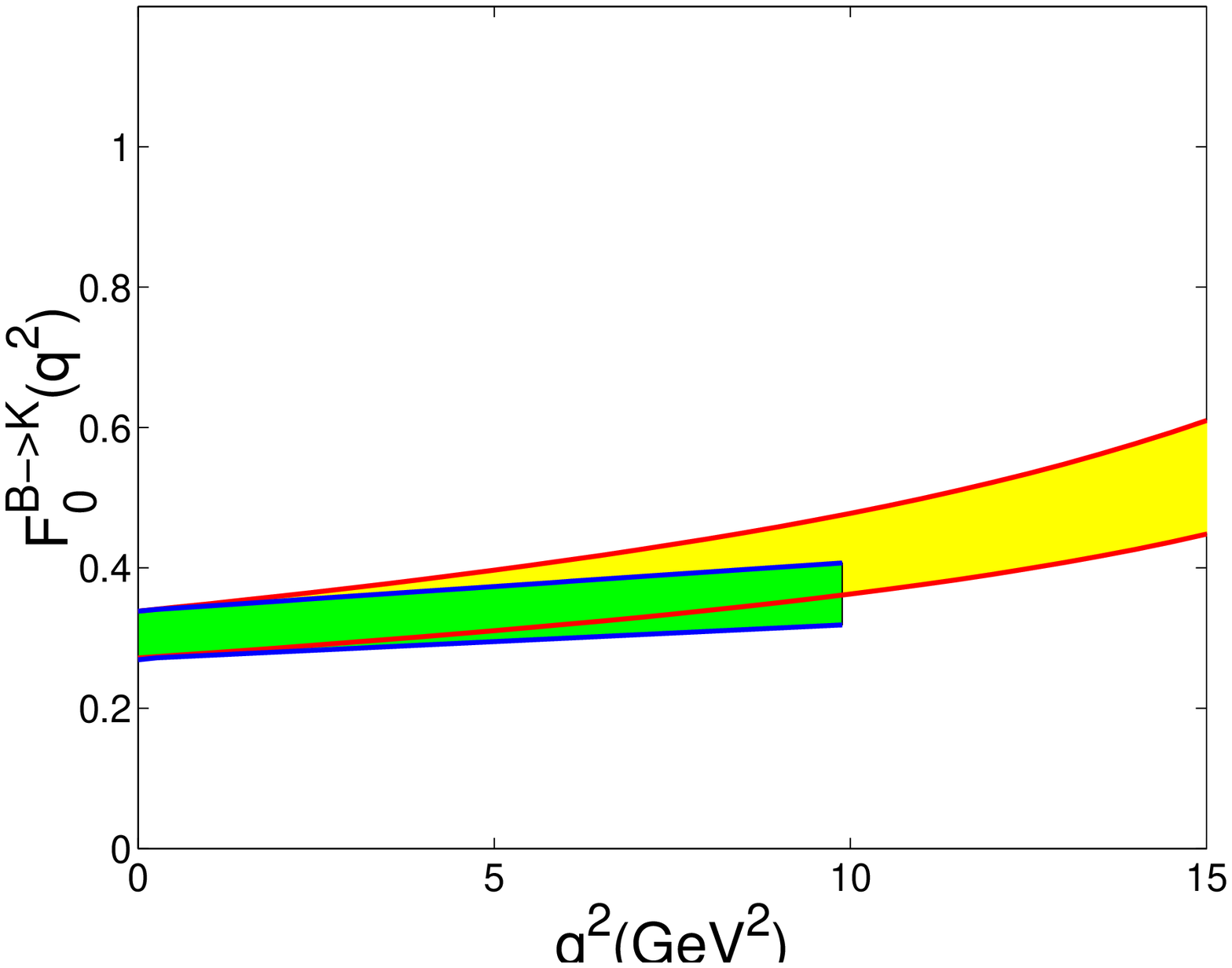} %
\includegraphics[width=0.3\textwidth]{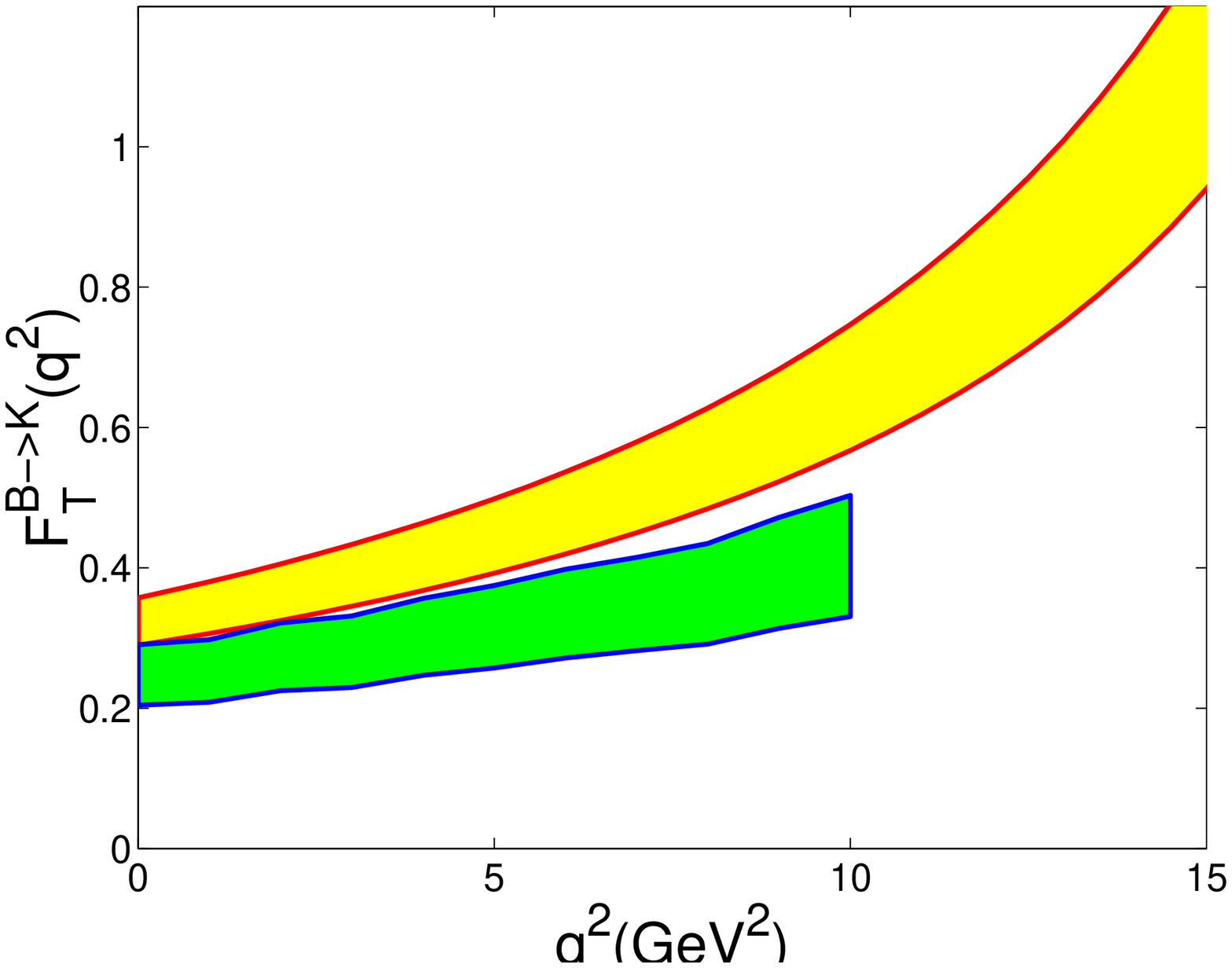}
\caption{Comparative results of $F_{+,0,T}^{B\to K}(q^2)$ within the
$k_T$ factorization approach and the QCD LCSR, where the fuscous
shaded band stands for $k_T$ factorization results and the fainter
band stands for the QCD LCSR results \cite{sumrule}. The upper edge
of the fuscous shaded band is for $\bar\Lambda=0.50GeV$ and
$\delta=0.30$, and the lower edge of the fuscous shaded band is for
$\bar\Lambda=0.55GeV$ and $\delta=0.25$. } \label{btok}
\end{figure}

The $B\to K$ transition form factors $F^{B\to K}_{+,0,T}$ are shown
in Fig.(\ref{btok}), where the results of the $k_T$ factorization
with $\bar\Lambda\in [0.50,0.55]$ and $\delta\in[0.25,0.30]$ are
shown by a fuscous shaded band with $0\leq q^2\leq 10GeV^2$ and the
results of the LCSR with its $(12\%+3\%)$ uncertainty \cite{sumrule}
are shown by a fainter band with $0\leq q^2\leq15GeV^2$, where the
extra $3\%$ error is introduced due to the $a^K_1(1GeV)$
uncertainty. The first Gegenbauer moment $a_1^K$ stands for the
$SU_f(3)$-breaking of the $B\to K$ transition form factors in
comparison to the $B\to \pi$ form factors, and it has been studied
by the light-front quark model \cite{quark1}, the QCD sum rule
\cite{lcsr1,pballa1k,ballmoments,Brauna1k} and the lattice
calculation \cite{lattice1,lattice2} and etc. In Ref.\cite{lcsr1},
the QCD sum rule for the diagonal correlation function of local and
nonlocal axial-vector currents is used, in which the contributions
of condensates up to dimension six and the ${\cal
O}(\alpha_s)$-corrections to the quark-condensate term are taken
into account. The moments derived there are close to that of the
lattice calculation \cite{lattice1,lattice2}, so we shall take
$a^K_1(1{\rm GeV})=0.05\pm 0.02$ and $a^K_2(1{\rm GeV})=0.10\pm
0.05$ to do our estimation. Further more a discussion of the
uncertainty of $a^K_1$ to the $k_T$ factorization approach can be
found in Ref.\cite{whf}, which shows that such kind of uncertainty
is quite small in comparison to the uncertainties caused by the
change of $\bar\Lambda$ and $\delta$.

Similar to the $B\to\pi$ case, it can be found that the form factors
$F^{B\to K}_{+,0,T}$ decrease with the increment of $\bar\Lambda$
and increase with the increment of $\delta$. The upper edge of the
fuscous shaded band is for $\bar\Lambda=0.50GeV$ and $\delta=0.30$,
and the lower edge of the fuscous shaded band is for
$\bar\Lambda=0.55GeV$ and $\delta=0.25$. One may observe that $k_T$
factorization results of $F^{B\to K}_{+,0}(q^2)$ can agree with that
of the QCD LCSR at small value of $q^{2}$ with proper values for
$\bar\Lambda$ and $\delta$. However the value of $F^{B\to
K}_{T}(q^2)$ is lower than that of QCD LCSR almost in the whole
$q^2$ region, so no proper ranges for $\bar\Lambda$ and $\delta$ can
be derived from the comparison of $F^{B\to K}_{T}(q^2)$. More
explicitly, at the largest recoil region $q^2=0$, the QCD LCSR gives
\cite{sumrule}: $F^{B\to K}_{+,0}(0)=0.331\pm
0.041+0.25(a^K_1(1GeV)-0.17)$ and $F^{B\to K}_{T}(0)=0.358\pm
0.037+0.25(a^K_1(1GeV)-0.17)$, which shows that $F^{B\to K}_{+,0}$
is {\bf smaller} than $F^{B\to K}_{T}(0)$ under the same parameters.
While the $k_T$ factorization approach gives: $F^{B\to
K}_{+,0}(0)=0.30\pm 0.04$ and $F^{B\to K}_{T}(0)=0.25\pm 0.03$ for
$\bar\Lambda\in[0.50,0.55]GeV$ and $\delta\in[0.25,0.30]$, which
shows that $F^{B\to K}_{+,0}$ is {\bf bigger} than $F^{B\to
K}_{T}(0)$ under the same parameters. This difference between the
$B\to\pi$ and $B\to K$ maybe explained by the $SU_f(3)$-breaking
symmetry effect. It can be found that within the $k_T$ factorization
and Ref.\cite{bsidepik}, a similar $SU_f(3)$-breaking effects can be
found for all the three form factors $F^{B\to K}_{+,0,T}(0)$, more
explicitly, $[F^{B\to K}_{+,0,T}(0)/F^{B\to\pi}_{+,0,T}(0)]\sim
1.08$ for the $k_T$ factorization approach and $[F^{B\to
K}_{+,0,T}(0)/F^{B\to\pi}_{+,0,T}(0)]\sim 1.24$ for
Ref.\cite{bsidepik}. While Ref.\cite{sumrule} gives somewhat
different $SU_f(3)$-breaking effects: $[F^{B\to
K}_{+,0}(0)/F^{B\to\pi}_{+,0}(0)] \sim 1.16$ and $[F^{B\to
K}_{T}(0)/F^{B\to\pi}_{T}(0)] \sim 1.28$. In the literature, new QCD
sum rules for $B\to \pi$, $K$ form factors have been derived from
the correlation functions expanded near the light cone in terms of
$B$-meson distribution, which are consistent with the present $k_T$
factorization results and show that $F^{B\to \pi}_{+,0}(0)=0.25\pm
0.05$, $F^{B\to \pi}_{T}(0)=0.21\pm 0.04$, $F^{B\to
K}_{+,0}(0)=0.31\pm 0.04$ and $F^{B\to K}_{T}(0)=0.27\pm 0.04$
\cite{bsidepik}. Very recently, another independent LCSR calculation
on the $B\to\pi$ and $B\to K$ transition form factors have been
presented in Refs.\cite{melic1,melic2}, which gives $F^{B\to
\pi}_{+,0}(0)=0.26^{+0.04}_{-0.03}$, $F^{B\to \pi}_{T}(0)=0.255\pm
0.035$, $F^{B\to K}_{+,0}(0)=0.36^{+0.05}_{-0.04}$ and $F^{B\to
K}_{T}(0)=0.38\pm 0.05$. Such results of $B\to\pi$ is very close to
our present PQCD results, while the results of $B\to K$ form factors
$F^{B\to K}_{+,0}(0)$ and $F^{B\to K}_{T}(0)$ are larger that ours,
and hence a larger $SU_f(3)$-breaking effect $[F^{B\to
K}_{T}(0)/F^{B\to\pi}_{T}(0)] \sim 1.49$ \cite{melic2} is presented
there. Such a discrepancy is mainly caused by a larger value
$a^K_1(1GeV)=0.10\pm 0.04$, the treatment of the $b$-quark mass and
also some other parameters like $f_B$ and etc.. So a comparative
study of the form factors and a precise QCD LCSR calculation on the
form factor with tensor current, including the $SU_f(3)$-breaking
effect and all its possible uncertainties, is necessary to clarify
the present situation \cite{whf1}. So we shall only make a
comparison of $F^{B\to K}_{+,0}(0)$ to decide possible range for
$\bar\Lambda$ and $\delta$ at the present. And by sampling 10,000
points for $F^{B\to K}_{+,0}(0)$ to be within the region derived
from QCD LCSR \cite{sumrule}, it can be found that all the points in
$\bar\Lambda\in[0.50,0.55]GeV$ and $\delta\in[0.25,0.30]$ are
allowable.

In the present paper, the properties of the B-meson light-cone wave
function up to next-to-leading order Fock state expansion have been
studied through a comparative study of the $B \to \pi$ and $B\to K$
transition form factors under both the $k_{T}$ factorization
approach and the QCD LCSR approach. The QCD LCSR approach with
proper correlator shall have no relation to the B-meson DA but shall
be quite sensitive to the light mesons' DAs, while the $k_{T}$
factorization approach is insensitive to the light mesons'
distribution amplitudes but depends on the B-meson DA heavily, so
these two approaches are compensated to each other. A more precise
QCD LCSR results shall be helpful to obtain a more accurate
information on the B-meson wave function, and vice versa.

We have applied the $k_{T}$ factorization approach to do a
systematical study on the $B \to \pi$ and $B\to K$ transition form
factors up to ${\cal O}(1/m_b^2)$, where the transverse momentum
dependence for the wave function, the Sudakov effects and the
threshold effects are included to regulate the endpoint singularity
and to acquire a more reasonable result. By comparing with the QCD
LCSR results, it has been found that when the two typical
phenomenological parameters $\bar\Lambda\in [0.50,0.55]$ and
$\delta\in[0.25,0.30]$ (the correlation relation between
$\bar\Lambda$ and $\delta$ can be found in the Right diagram of
Fig.(\ref{pirange})), which control the leading and next-to-leading
Fock states' contributions respectively, the results of
$F^{B\to\pi}_{+,0,T}(q^2)$ and $F^{B\to K}_{+,0,T}(q^2)$ from these
two approaches are consistent with each other in the large recoil
energy region. Inversely, one can derive the reasonable regions for
the two undetermined parameters of the simple B-meson model wave
function as shown in Eqs.(\ref{bw1}, \ref{bw2}). The slight
discrepancy ($\sim 20\%$) of $F^{B\to\pi,K}_{T}(q^2)$ between the
$k_T$ factorization approach and the QCD LCSR results of
Ref.\cite{sumrule} may be compensated by carefully taking the
$SU_f(3)$-breaking effects into the QCD sum rule calculation. As a
byproduct, it can be found that the $SU_f(3)$-breaking effects are
small in the $B\to K$ transition form factors. Finally, one can
adopt the B-meson wave functions up to next-to-leading order Fock
state expansion to present a more precise studies on the B-meson
decays up to ${\cal O}(1/m_b^2)$. It is noted that at the present,
only the main uncertain sources are considered to determine the
properties of the B-meson wave function, a more precise study that
includes a more precise pseudo-scalar wave functions shall be
helpful to improve our understanding on the B-meson wave function.

\begin{center}
\section*{Acknowledgements}
\end{center}

The authors would like to thanks B. Melic for helpful discussions on
the QCD LCSR analysis of the form factors. This work was supported
in part by Natural Science Foundation Project of CQ CSTC under grant
number 2008BB0298 and Natural Science Foundation of China under
grant number 10805082, and by the grant from the Chinese Academy of
Engineering Physics under the grant numbers: 2008T0401
and 2008T0402.\\

\end{document}